\documentclass[aip,jap,jmp,amsmath,amssymb,reprint,floatfix]{revtex4-1} 
\usepackage[utf8x]{inputenc}
\usepackage{graphicx}
\usepackage{dcolumn}
\usepackage[version=3]{mhchem}
\usepackage{units}
\usepackage{hyperref}

\begin{document}
\hypersetup{pdfauthor={D. Weber and U. Poppe},pdftitle={Non-volatile gated variable resistor based on doped La2CuO4 and SrTiO3 heterostructures}}


\title{Non-volatile gated variable resistor based on doped \ce{La2CuO_{4+$\delta$}} and \ce{SrTiO3} heterostructures} 



\author{Dieter Weber}
\email[]{d.weber@fz-juelich.de}

\author{Ulrich Poppe}
\email[]{u.poppe@fz-juelich.de}

\affiliation{Peter Gr\"{u}nberg Institute PGI-5, Forschungszentrum J\"{u}lich, D-52425 J\"{u}lich, Germany}


\date{2 March 2012}

\begin{abstract}
Gated variable resistors were manufactured by depositing epitaxial heterostructures of doped \ce{La2CuO_{4+$\delta$}} and \ce{SrTiO3} layers. Their conductance change as function of write current $I$ and write time $t$ followed a simple empirical law of the form $\Delta G/G = C I^A t^B$. This behavior is in agreement with ionic transport that accelerates exponentially with electrical field strength.
\end{abstract}


\maketitle 
Solid-state ionic transport on the nanoscale has recently attracted attention because it is believed to be crucial for the operation of two possible alternative concepts for computer memory: electrochemical metallization cells and devices based on local valence change of the cations through anion migration.\cite{Waser2007} Although the applied voltages are only moderate, there are strong electrical fields present that accelerate the ionic transport exponentially by contributing part of the activation energy for ionic transport\cite{Lamb-field-acceleration, Meyer-CMOX, Strukov2009}. This contrasts with conventional ionic transport where the activation energy is solely provided by thermal excitation. Measuring the ionic transport in nanometer-thick films can be challenging, especially in the case of oxygen.
\begin{figure}
 \centering
 \includegraphics[width=7.5cm]{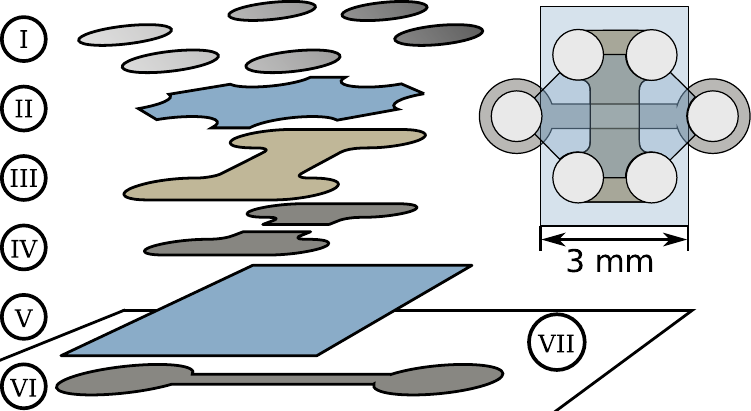}
 \caption{Sample structure from top to bottom. (I) silver contact pads; (II) passivation; (III) channel; (IV) channel contacts; (V) dielectric and ionic conductor; (VI) gate; (VII) substrate. The junction area is about 0.5\,mm²}
 \label{fig:Masks}
\end{figure}

We created a device to measure the change of in-plane conductance caused by the doping effect of out-of-plane oxygen anion transport similar to one reported by Ramesham et al.\cite{ramesham:1935}. Epitaxial oxide structures were deposited through titanium shadow masks with high-pressure pure oxygen sputtering at 800\,°C (heater temperature) on epi-polished single-crystal \ce{SrTiO3} (100)K substrates from CrysTec (Berlin). A sketch of the different layers is shown in Fig.~\ref{fig:Masks} and the materials and thicknesses are given in Tab.~\ref{tab:conditions}. A sputtered silver layer of 600\,nm thickness was used for the electrical connections to the oxide films.
\begin{table}
\caption{Layer overview. The given thicknesses are approximate peak values for a typical sample in the middle of the mound-shaped structures.}
\label{tab:conditions}
\begin{tabular}{llrc}
\toprule
\textbf{Part} & \textbf{Material} & \textbf{d/nm} & \textbf{Doping}\\
\colrule
Gate             & \ce{La_{1.85}Sr_{0.15}CuO_{4+$\delta$}} & 20 & p \\
Dielectric       & \ce{SrTiO3}                 & 25 & intrinsic\\
Channel contacts & \ce{La_{1.85}Sr_{0.15}CuO_{4+$\delta$}} & 20 & p\\
Channel          & \ce{La2CuO_{4+$\delta$}}                & 5  & p\\
Passivation      & \ce{SrTiO3}                 & 10 & intrinsic\\
\botrule
\end{tabular}
\end{table}

For each layer, the chamber was first pumped down to a background pressure of less than $\unit[1\times10^{-5}]{mbar}$. Then the sample was heated to the deposition temperature in an atmosphere of 0.5\,mbar oxygen within 45 minutes. Afterwards the sputter target was pre-sputtered for one hour, followed by layer deposition lasting from 30 minutes to 3 hours, depending on deposition rate and desired film thickness. The sample was finally cooled down in an atmosphere of 0.5\,mbar oxygen. 

The deposition rate was determined by measuring the thickness of the device structures with a profilometer after a very long deposition time to achieve sufficient film thickness and by examining cross-sections with SEM and TEM. The layer thickness is not uniform over the device area due to the shadowing effect of the masks that leads to very shallow slopes on the edges. The given thicknesses are therefore to be understood as approximate peak values in the middle of the structures.

\ce{La2CuO_{4+$\delta$}} was chosen as the channel material because excess oxygen intercalates easily into the crystal structure even at room temperature and changes the material's behavior from semiconducting to metallic at sufficient concentration.\cite{Daridon1997118}

Electrical properties were measured with a Keithley 6221 AC and DC current source, coupled with a Keithley 2182A nanovoltmeter. The positive current direction through the dielectric is defined as from the channel to the gate.
\begin{figure}
 \centering
  \includegraphics[width=8.5cm]{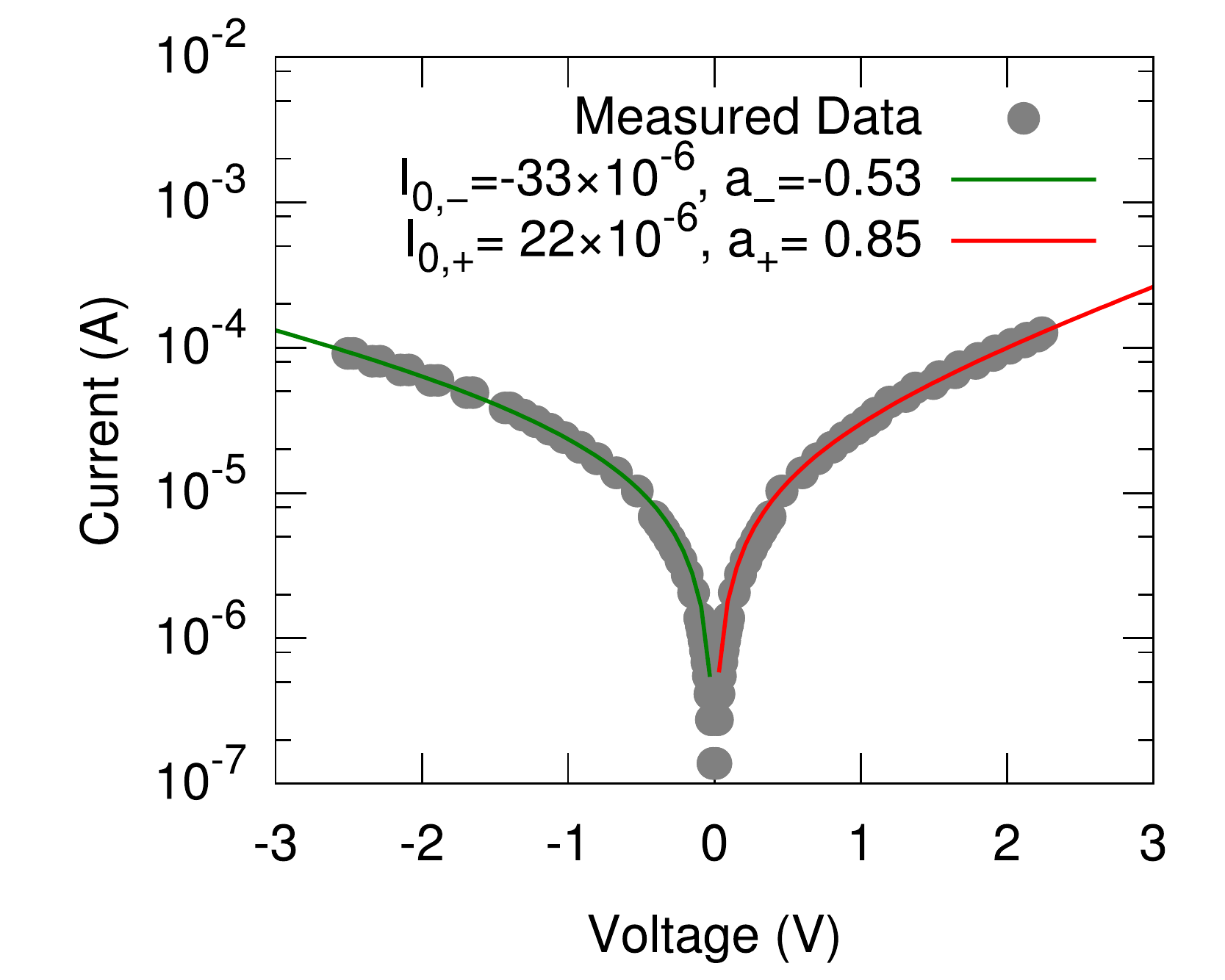}
 \caption{Current-voltage curve of the dielectric. Points: measurement; solid lines: fit with the indicated parameters for Eq.~\eqref{eq:I_el}}
 \label{fig:I-U-Dielectric}
\end{figure}

The electronic current $I_{el}$ through the dielectric as a function of voltage $U$ can be described by the equation
\begin{equation}
  I_{el} = I_{0,el} \left( \exp \left( a_{el} \times U \right) - 1 \right) \label{eq:I_el}
\end{equation}
with fit parameters $I_{0, el}, a_{el}$ being different for the positive and negative direction (Fig.~\ref{fig:I-U-Dielectric}). This equation corresponds to the Shockley equation that describes the behavior of a diode. Yajima et al.\cite{Yajima2011} obtained comparable results with Al/\ce{SrTiO3}/\ce{La_{0.7}Sr_{0.3}MnO3 }/Nb:\ce{SrTiO3} heterojunctions. Hikita et al. \cite{PhysRevB.79.073101} also reported that the barrier height of \ce{La_{0.7}Sr_{0.3}MnO3 }/Nb:\ce{SrTiO3} depends sensitively on the surface termination of the interface. This strong influence of termination could explain the spread of one order of magnitude in dielectric properties observed by us between two devices manufactured in one process on the same substrate.

\begin{figure}
 \centering
 \includegraphics[width=8.5cm]{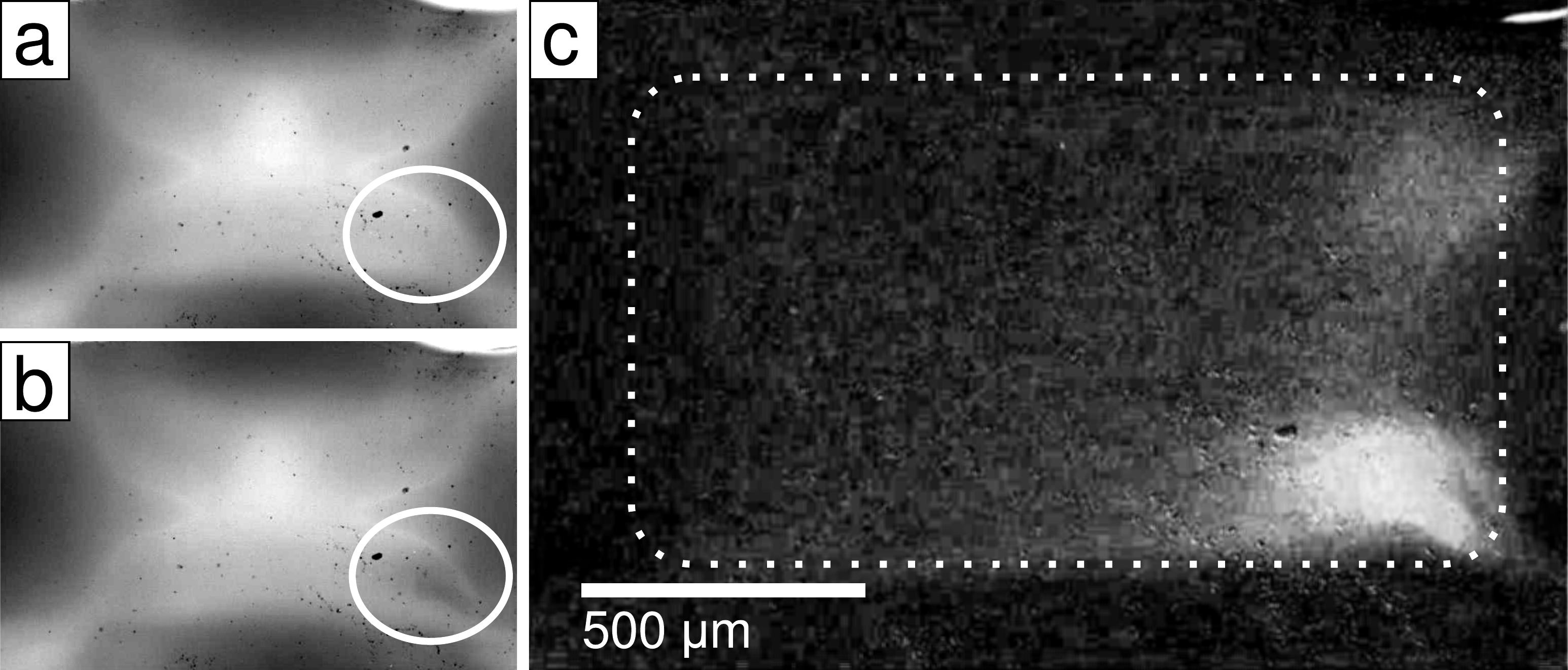}
 \caption{Reversible changes visible in an optical microscope (reflected light) after writing for 10~minutes with +20\,mA. (a) and (b) show the sample before and after writing, respectively, while (c) is the difference of the two images. The dotted line in (c) outlines the crossing area between channel and gate (compare Fig.~\ref{fig:Masks}). The current flowed between the lower (channel) and the right (gate) side, and the change is localized in the area of maximum field strength in the dielectric.}
 \label{fig:visible}
\end{figure}

The current-voltage characteristics of the dielectric could only be measured accurately under conditions where the voltage drop over channel and gate, forming the electrodes (resistance typically between 1\,kΩ and 10\,kΩ), was small compared to the voltage drop across the dielectric. Only under such conditions is the voltage drop across the dielectric nearly constant over the entire electrode area and can be measured in a four-point configuration using the two channel and gate electrodes. In other cases, the voltage also drops significantly inside the large-area electrodes, which distorts the measurements especially at higher currents, as the current through the dielectric grows exponentially with applied voltage, while the electrodes show ohmic behavior. This effect is responsible for the localized writing effect in Fig.~\ref{fig:visible}, where the visible change is concentrated in the corner with maximum voltage drop across the dielectric.
\begin{figure}
 \centering
 \includegraphics[width=8.5cm]{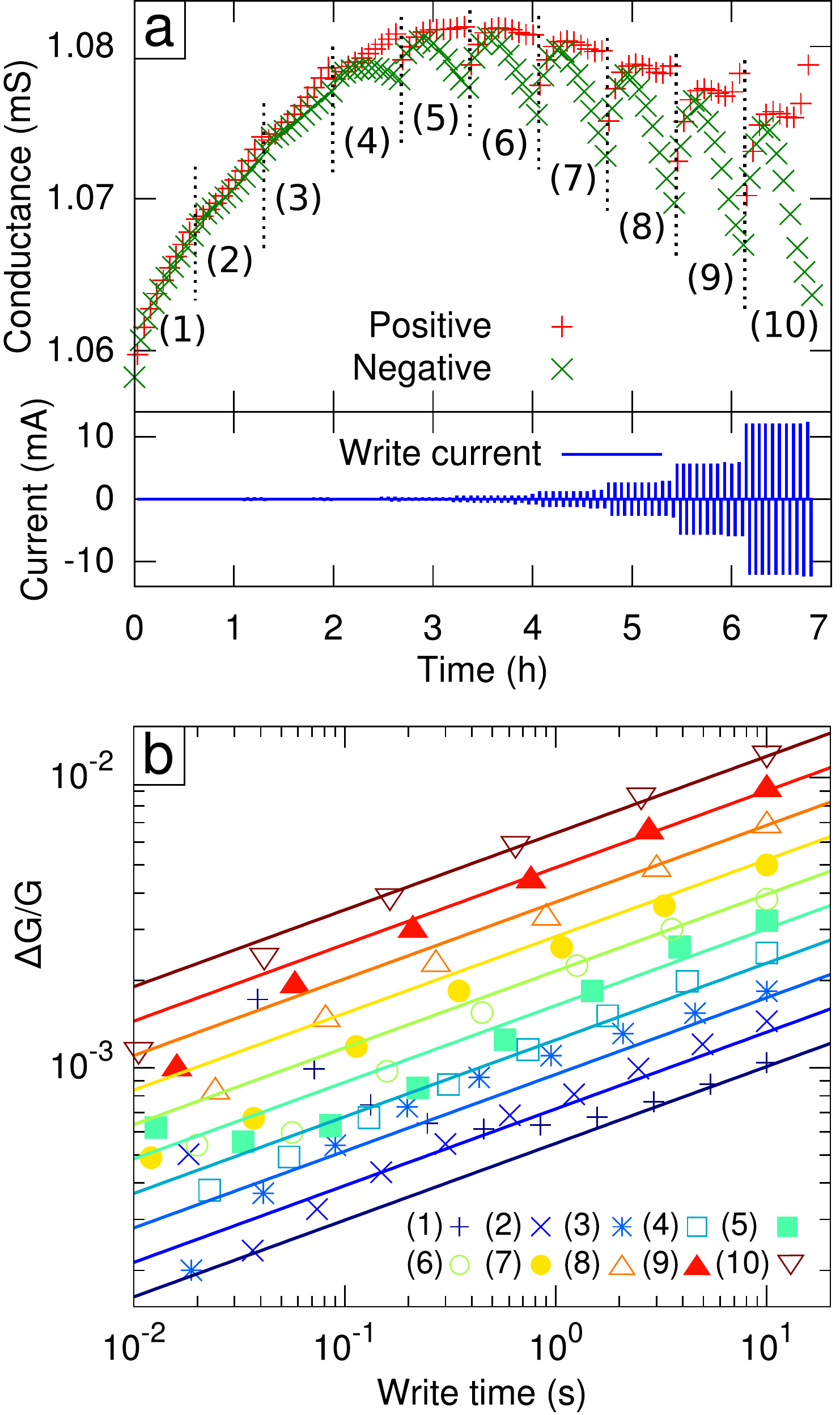}
 \caption{Conductance change during a writing experiment with increasing current. (a) shows the the absolute conductance values and write pulses over the course of the experiment, while in (b) the conductance change relative to the previous measured point is plotted as function of write time for positive write currents, along with a fit of Eq.~\ref{eq:DG_G} ($C$=0.032, $A$=0.36, $B$=0.27) as solid lines. Data below 10\,ms write time is omitted because of dominating noise and drift. The corresponding plot for negative currents would be similar. Legend (write current, minimum write time): (1) 13.8\,µA, 38.6\,ms; (2) 29.2\,µA, 18.2\,ms; (3) 61.2\,µA, 8.55\,ms; (4) 132\,µA, 4.03\,ms; (5) 280\,µA, 1.89\,ms; (6) 596\,µA, 892\,µs; (7) 1.27\,mA, 420\,µs; (8) 2.69\,mA, 198\,µs; (9) 5.71\,mA, 93.0\,µs; (10) 12.1\,mA, 43.8\,µs}
 \label{fig:RW}
\end{figure}

To investigate the switching behavior of the channel, the device was written with current pulses of alternating direction, varied duration and increasing magnitude through the dielectric (compare lower part of Fig.~\ref{fig:RW}a). For each current, a minimum write time was chosen to transport enough charge to reach the necessary write voltage, while the maximum write time was always 10\,s. The conductance of channel and gate was measured between write pulses. A current from the channel to the gate increases the conductance of the channel (Fig.~\ref{fig:RW}) and, to a slightly lesser degree, decreases the conductance of the gate, and vice versa. The relative conductance change $\frac{\Delta G}{G}$ as a function of write current $I$ and write time $t$ follows the empirical law
\begin{equation}
 \frac{\Delta G}{G} = \left(\frac{I}{I_{0,ion}}\right)^A \left(\frac{t}{t_0}\right)^B = C I^A t^B \label{eq:DG_G}
\end{equation}
with fit parameters $C = (1/I_{0,ion})^A(1/t_0)^B$, $A$ and $B$ over a time range of about three orders of magnitude and a current range of about two orders of magnitude (Fig.~\ref{fig:RW}b). All three parameters differed from sample to sample, with $C$ being between 0.01 and 0.2, and $A$ and $B$ between 0.2 and 0.8. The data are less accurate for weak currents and short write times due to drifting over the course of the experiment and the charge lost in parasitic capacitances. The time dependence was further investigated with write pulses of up to one hour and up to 20\,mA. Pulses of 10\,min length lead to reversible visible changes (Fig.~\ref{fig:visible}), and longer pulses degrade the sample more and more under bubble formation. Very strong writing decreases the conductance irreversibly by more than one order of magnitude, while the the $t^A$ time-dependence law is still obeyed. The conductance drifts back towards higher conductance after writing to a low-conductance state with a similar dependence on time over at least one hour.  We also found a power law time dependence in the data of Ramesham et al.\cite{ramesham:1935}: Their data for channel conductance versus time of their \ce{WO3}/\ce{Cr2O3} device, measured with two different positive gate biases, can be fitted with the formula $G(t) = 2/3 \times 10^{-11} \times t^{1.5}$ (+20\,V) resp. $G(t) = 2 \times 10^{-10} \times t^{1.7}$ (+40\,V). Such a behavior is also observed in the relaxation of many dielectrics, caused by slowly moving charges like hopping electrons or ions\cite{0022-3727-32-14-201}. Ang et al.\cite{ang:818} specificly observed such a behavior in \ce{SrTiO3} that is stressed under high voltage.

The current dependence (Eq.~\ref{eq:DG_G}) can be transformed to a voltage dependence using the parameters from Eq.~\ref{eq:I_el} if the exponential term dominates over the ``$-1$'' term, leading to
\begin{equation}
\frac{\Delta G}{G} = \left(\frac{I_{0,el}}{I_{0,ion}}\right)^A \exp \left(a_{el}\times A\times U \right) \left(\frac{t}{t_0}\right)^B,
\end{equation}
which is in agreement with an exponential increase of the ionic transport with field strength\cite{Lamb-field-acceleration, Meyer-CMOX}.

In conclusion, we present a non-volatile variable resistor based on epitaxial oxide heterostructures. The resistance change as a function of write current and write time follows a simple empirical law over a wide range of currents and times. The time-domain behavior is similar to dielectric relaxation caused by slowly mobile charge carriers. The current dependence in connection with bubble formation can be interpreted as evidence for ionic transport that accelerates exponentially with increasing field strength and is very slow in absence of a field that contributes to the activation energy for transport. Exponential increase of write speed with voltage was observed perviously by Terabe et~al. for variable metal filaments\cite{Terabe2005} and by Pickett et~al. for tunneling contacts\cite{Pickett2009}. Measuring the conductance change of a doped semiconductor allows more direct conclusions on the amount of transported ions: The conductance change should be proportional to the change in carrier density and mobility and therefore proportional to the amount of transported ions in the range of small relative concentration changes. Our device can therefore be a valuable tool for studying low-level oxygen transport parallel to a large electronic current in nanometer-thick films under strong electrical fields over a wide parameter range. A gated variable resistor also has potential applications in reconfigurable logic circuits or artificial neuronal networks.\cite{ramesham:1935,1609383,sakamoto:252104,APEX.4.015204} 

\end{document}